\documentclass[preprint,nofootinbib]{revtex4}
\usepackage{ulem}
\usepackage{floatrow}
\usepackage[font=small,labelfont=bf,tableposition=top]{caption}
\usepackage{graphicx}
\usepackage{epstopdf}
\usepackage[colorlinks=true,linkcolor=red]{hyperref}
\usepackage{bm}
\usepackage{booktabs,tabularx}
\usepackage{floatrow}
\usepackage{threeparttable}
\usepackage{mathrsfs}
\usepackage{setspace}
\newcommand{\bea}{\begin{eqnarray}}
\newcommand{\eea}{\end{eqnarray}}
\newcommand{\beq}{\begin{equation}}
\newcommand{\eeq}{\end{equation}}
\newcommand{\nn}{\nonumber}
\def\/{\over}

\begin{document}

\title{Interaction between Unruh-Dewitt detectors exclusively due to acceleration: A Parallel to the FDU Effect}

\author{Wenting Zhou$^{1,2}$, Shijing Cheng$^{3}$, and Hongwei Yu$^{4,}$\footnote{Corresponding author: hwyu@hunnu.edu.cn}}
\affiliation{
$^{1}$ Institute of Fundamental Physics and Quantum Technology, Ningbo University, Ningbo, Zhejiang 315211, China\\
$^{2}$ Department of Physics, School of Physical Science and Technology, Ningbo University, Ningbo, Zhejiang 315211, China\\
$^{3}$ School of Physics and Information Engineering, Shanxi Normal University, Taiyuan, 030031, China\\
$^{4}$ Department of Physics, Synergetic Innovation Center for Quantum Effect and Applications and Institute of Interdisciplinary Studies, Hunan Normal University, Changsha, Hunan 410081, China}

\begin{abstract}

We have discovered an interaction between two detectors in a vacuum that emerges exclusively due to acceleration, akin to  the spontaneous excitation of a single detector as predicted by the Fulling-Davies-Unruh (FDU) effect. However, this interaction contrasts sharply with the FDU effect, which suggests that a uniformly accelerated detector behaves as if it were in a thermal bath, as the discovered interaction does not manifest in a thermal environment. The novel interaction displays unique dependencies on the separation between detectors: it can be either attractive or repulsive, with the potential to transition between these behaviors as the inter-detector separation changes. More intriguingly, it exhibits a surprising large-small duality in its dependence on acceleration, suggesting the existence of an optimal acceleration at which the interaction is strongest, in contrast to the monotonic acceleration-dependence of the FDU effect.

\end{abstract}
\maketitle

\section{Introduction}
The past decades have witnessed remarkable advancements in quantum field theory in curved spacetimes, with the Fulling-Davies-Unruh (FDU) effect~\cite{Fulling73,Davies75,Unruh76} standing out as one of the most significant phenomena. The FDU effect posits that a ground-state Unruh-DeWitt detector undergoes spontaneous excitation when uniformly accelerated in a vacuum, as if it were immersed in a thermal bath at a temperature proportional to its proper acceleration $a$. 
This effect underscores the observer-dependent nature of the quantum vacuum and particles, playing a crucial role in our understanding of quantum fields across various contexts. Furthermore, the FDU effect is closely related to other quantum phenomena in gravitational backgrounds, such as Hawking radiation from black holes~\cite{Unruh76} and the Gibbons-Hawking effect in de Sitter spacetime~\cite{Gibbons77,Deser97}.

Since the seminal works~\cite{Fulling73,Davies75,Unruh76}, the FDU effect has spurred significant interest in acceleration-dependent phenomena~\cite{Audretsch95,Passante98,Rizzuto07,Zhu10,Martin11,Hu12,Quach22,Kok03,Nesterov20,Benatti04,Zhang2007,Lin2008,Lin10,Mann2015,Ostapchuk12,Hu15,Yang16,BYZhou21,Koga19,ZHLiu21}.  These studies often deal with acceleration-induced modifications to quantum phenomena already present in the inertial case, such as energy shifts~\cite{Audretsch95,Passante98}, the Casimir-Polder force~\cite{Rizzuto07,Zhu10}, the Berry phase~\cite{Martin11,Hu12,Quach22}, quantum decoherence~\cite{Kok03,Nesterov20}, and quantum entanglement~\cite{Benatti04,Zhang2007,Lin2008,Mann2015,Lin10,Ostapchuk12,Hu15,Yang16,BYZhou21,Koga19,ZHLiu21}. However, these effects fundamentally differ from the FDU effect, which is a unique quantum effect solely caused by acceleration, absent in the inertial case.

The FDU effect is a novel quantum phenomenon associated with the radiative properties of a single detector. 
When two detectors are involved, new quantum effects emerge. One remarkable example is the dispersion interaction that arises due to the presence of vacuum fluctuations of quantum fields, which induce instantaneous dipole moments in the detectors and thus generate inter-detector interaction. Based on the FDU effect, one might expect that acceleration would only induce  modifications superimposed on the interaction between two inertial detectors.

Contrary to this expectation, we reveal in this letter an interaction between two detectors that exclusively arises from acceleration. Our finding stems from an investigation of
the interaction between two electrically polarizable detectors interacting with fluctuating electromagnetic fields in a vacuum. We demonstrate that a novel interaction emerges between cross-polarizable detectors only  when they are uniformly accelerated. Here, `cross-polarizable' means that the two detectors are polarizable along the inter-detector separation and a perpendicular direction, respectively. This interaction is extraordinary, as no such interaction exists between these detectors when they are inertial, either in a vacuum or in a thermal bath, a fact that, although less acknowledged, has been implied in the literature (see, for instance, Eq. (50) of
Ref.~\cite{Casimir-Polder48} for the vacuum case and Eqs.~(50), and (53)-(55) in Ref.~\cite{Cheng23} for the thermal one).

The interaction we discovered is a quantum effect arising exclusively due to acceleration, on par with the FDU effect, as it implies that  the interaction emerges only when the detectors are accelerated, much like how a single ground-state detector spontaneously excites only when it is accelerated. Notably, this interaction possesses several unique characteristics. First, it exhibits behaviors with respect to the inter-detector separation that are dramatically different from the usual dispersive interactions between non-cross-polarizable atoms. Second, this interaction vanishes not only when the acceleration becomes extremely small but also, surprisingly, when it tends to be infinitely large. Furthermore, it displays a large-small duality in its acceleration-dependence: it scales as $a^2$ when the acceleration $a$ is very small and as $a^{-2}$ when $a$ is very large. Third, this interaction can switch between attractive and repulsive as the inter-detector separation $L$ varies, particularly when $a$ and $L$ are comparable to the transition frequency and the wavelength of the detectors, respectively.

\section{Model and method}
Consider two Unruh-DeWitt detectors, $A$ and $B$, that are uniformly and synchronously accelerated in a vacuum along the following trajectories:
\bea
\mathrm{x}^A(\tau)=\left(a^{-1}\sinh{(a\tau)},a^{-1}\cosh{(a\tau)},0,0\right),\label{trajectoryA}\\
\mathrm{x}^B(\tau)=\left(a^{-1}\sinh{(a\tau)},a^{-1}\cosh{(a\tau)},0,L\right),\label{trajectoryB}
\eea
where $\tau$ denotes the proper time and $L$ is the inter-detector separation, and
$a$ is the uniform acceleration. The two detectors are modeled as two-level quantum systems, initially in their ground states, and they are cross-polarizable.  Specifically,  detector $A$ is polarizable along the inter-detector separation and detector $B$ is polarizable along the direction of acceleration. As a result, their dipole moments are represented by $\bm{\mu}^A=(0,0,\mu^A)$ and $\bm{\mu}^B=(\mu^B,0,0)$. The Hamiltonian of the `detectors+field' system takes the form
\beq
H(\tau)=\sum_{n,\xi}\omega^{\xi}_n\sigma_{nn}^{\xi}(\tau)
+\sum_{\mathbf{k},\lambda}\omega_{\mathbf{k}}a^{\dag}_{\mathbf{k},\lambda}(t)a_{\mathbf{k},\lambda}(t)\frac{dt}{d\tau}
-\bm{\mu}^A\cdot\mathbf{E}(\mathrm{x}_A(\tau))-\bm{\mu}^B\cdot\mathbf{E}(\mathrm{x}_B(\tau))\;,
\label{total}
\eeq
where $n=g$ or $e$ labels the ground or the excited state of a single detector, $\omega^{\xi}_n$ denotes the energy of detector $\xi(=A,B)$ in the state $|n\rangle$, $\sigma_{nn}=|n\rangle\langle n|$, $\nu$ and $\mathbf{k}$ are the polarization index and the wave vector of the electromagnetic field, and $a^{\dag}_{\mathbf{k},\nu}(t)$ and $a_{\mathbf{k},\nu}(t)$ are the creation and annihilation operators of the electric field $\bf{E}(x(\tau))$, respectively. As mentioned in the introduction, no interaction exists between two cross-polarizable detectors when $a=0$ whether in a vacuum or a thermal bath.  However,  what happens when acceleration is present ($a\neq0$)?

We assume that the detector-field coupling is weak, allowing us to use a perturbative approach to calculate the inter-detector interaction. We will utilize the formalism proposed by Dalibard, Dupont-Roc, and Cohen-Tannoudji (the DDC formalism)~\cite{DDC82,DDC84}. Originally established for second-order calculations, this formalism has been widely applied to explore various second-order quantum effects arising from the interaction between a small quantum system $\mathcal{J}$ and a large reservoir $\mathcal{R}$~\cite{Meschede90,Rizzuto07,Zhu10,Audretsch94,Audretsch95,Passante98,Tomazelli03,Zhou07}. Compared to the standard perturbative method, the DDC formalism offers an alternative perspective by interpreting quantum phenomena in terms of the evolution of observables of $\mathcal{J}$, separately accounting for the contributions of field fluctuations from $\mathcal{R}$ and the radiation reaction of $\mathcal{J}$. Recently, this formalism has been extended from second-order to fourth-order calculations~\cite{Passante14,Cheng22,Zhou21,Cheng23}.

Following similar procedures as outlined in Ref.~\cite{Cheng23}, we find that the contribution of vacuum fluctuations [vf-contribution] to the interaction energy of two detectors synchronously moving in a vacuum can be expressed as
\bea
(\delta E)_{vf}&=&4i\int_{\tau_0}^{\tau}\mathrm{d}\tau_1\int_{\tau_0}^{\tau_1}\mathrm{d}\tau_2\int_{\tau_0}^{\tau_2}\mathrm{d}\tau_3\;C^F(\mathrm{x}_A(\tau),\mathrm{x}_B(\tau_3))
\chi^F(\mathrm{x}_A(\tau_1),\mathrm{x}_B(\tau_2))\nonumber\\&&\times
\chi^A(\tau,\tau_1)\chi^B(\tau_2,\tau_3)\;,
\label{vf-contriution}
\eea
where $\tau_0$ is the onset time of the detector-field interaction, $C^F(\mathrm{x}_A(\tau),\mathrm{x}_B(\tau'))$ and $\chi^F(\mathrm{x}_A(\tau),\mathrm{x}_B(\tau'))$, the symmetric correlation function and the linear susceptibility respectively, are defined as
\bea
C^F(\mathrm{x}_A(\tau),\mathrm{x}_B(\tau'))\equiv{1\/2}\langle0|\left\{E^f_z(\mathrm{x}_A(\tau)),E^f_x(\mathrm{x}_B(\tau'))\right\}|0\rangle\label{CF}\quad\quad\\
\chi^F(\mathrm{x}_A(\tau),\mathrm{x}_B(\tau'))\equiv{1\/2}\langle 0|\left[E^f_z(\mathrm{x}_A(\tau)),E^f_x(\mathrm{x}_B(\tau'))\right]|0\rangle
\theta(\tau-\tau')
\label{chiF}
\eea
with $E_z^f$ and $E_x^f$ denoting the $z$- and $x$-component of the free electric field and $|0\rangle$ representing the vacuum state respectively, and $\chi^{\xi}(\tau,\tau')$,  the antisymmetric statistical function of detector $\xi$, is given by
\bea
\chi^{\xi}(\tau,\tau')\equiv{1\/2}\langle g_{\xi}|\left[\mathbf{\mu}^{\xi,f}(\tau),\mathbf{\mu}^{\xi,f}(\tau')\right]|g_{\xi}\rangle
={1\/2}|\mu^{\xi}_{ge}|^2\left(e^{-i\omega_{\xi}(\tau-\tau')}-e^{i\omega_{\xi}(\tau-\tau')}\right)
\label{Chi-atom}
\eea
with $\mathbf{\mu}^{\xi,f}(\tau)$ 
being the free part of the dipole moment operator of detector $\xi$ and $\mu^{\xi}_{ge}=\langle g_{\xi}|\mu^{\xi,f}(\tau_0)|e_{\xi}\rangle$~\cite{Cheng23}. Similarly, the contributions of the radiation reaction of the detectors [rr-contribution] can be expressed as
\bea
&&(\delta E)_{rr}\nn\\
&=&4i\int_{\tau_0}^{\tau}\mathrm{d}\tau_1\int_{\tau_0}^{\tau_1}\mathrm{d}\tau_2\int_{\tau_0}^{\tau_2}\mathrm{d}\tau_3\;\chi^F(\mathrm{x}_A(\tau),\mathrm{x}_B(\tau_3))
\chi^F(\mathrm{x}_A(\tau_1),\mathrm{x}_B(\tau_2))C^A(\tau,\tau_1)\chi^B(\tau_2,\tau_3)\nonumber\\&&
+4i\int_{\tau_0}^{\tau}\mathrm{d}\tau_1\int_{\tau_0}^{\tau_1}\mathrm{d}\tau_2\int_{\tau_0}^{\tau_2}\mathrm{d}\tau_3\;\chi^F(\mathrm{x}_A(\tau_1),\mathrm{x}_B(\tau_3))
\chi^F(\mathrm{x}_A(\tau),\mathrm{x}_B(\tau_2))C^A(\tau,\tau_1)\chi^B(\tau_3,\tau_2)\nonumber\\&&
+4i\int_{\tau_0}^{\tau}\mathrm{d}\tau_1\int_{\tau_0}^{\tau_1}\mathrm{d}\tau_2\int_{\tau_0}^{\tau_2}\mathrm{d}\tau_3\;\chi^F(\mathrm{x}_B(\tau_2),\mathrm{x}_A(\tau_3))
\chi^F(\mathrm{x}_A(\tau),\mathrm{x}_B(\tau_1))C^A(\tau,\tau_3)\chi^B(\tau_1,\tau_2)\nonumber\\&&
+4i\int_{\tau_0}^{\tau}\mathrm{d}\tau_1\int_{\tau_0}^{\tau}\mathrm{d}\tau_2\int_{\tau_0}^{\tau_2}\mathrm{d}\tau_3\;\chi^F(\mathrm{x}_A(\tau_2),\mathrm{x}_B(\tau_3))
\chi^F(\mathrm{x}_A(\tau),\mathrm{x}_B(\tau_1))\chi^A(\tau,\tau_2)C^B(\tau_1,\tau_3)\nonumber\\&&
+4i\int_{\tau_0}^{\tau}\mathrm{d}\tau_1\int_{\tau_0}^{\tau_1}\mathrm{d}\tau_2\int_{\tau_0}^{\tau}\mathrm{d}\tau_3\;C^F(\mathrm{x}_A(\tau_3),\mathrm{x}_B(\tau_2))
\chi^F(\mathrm{x}_A(\tau),\mathrm{x}_B(\tau_1))\chi^A(\tau,\tau_3)\chi^B(\tau_1,\tau_2)\nonumber\\&&
+4i\int_{\tau_0}^{\tau}\mathrm{d}\tau_1\int_{\tau_0}^{\tau_1}\mathrm{d}\tau_2\int_{\tau_0}^{\tau_1}\mathrm{d}\tau_3\;\chi^F(\mathrm{x}_A(\tau),\mathrm{x}_B(\tau_3))
\chi^F(\mathrm{x}_A(\tau_1),\mathrm{x}_B(\tau_2))C^A(\tau,\tau_1)\chi^B(\tau_3,\tau_2),\;\;
\label{rr-contriution}
\eea
where
\bea
C^{\xi}(\tau,\tau')&\equiv&{1\/2}\langle g_{\xi}|\left\{\mathbf{\mu}^{\xi,f}(\tau),\mathbf{\mu}^{\xi,f}(\tau')\right\}|g_{\xi}\rangle
={1\/2}|\mu^{\xi}_{ge}|^2\left(e^{i\omega_{\xi}(\tau-\tau')}+e^{-i\omega_{\xi}(\tau-\tau')}\right)
\label{C-atom}
\eea
is the symmetric statistical function of detector $\xi$.

\section{Interaction energy of two detectors uniformly accelerated in vacuum.}
Based on the formulae above, we derive the following interaction energy for the two detectors moving along the trajectories described by Eqs. (\ref{trajectoryA}) and (\ref{trajectoryB}) (the detailed derivation is provided in Appendix~\ref{Appendix-derivation1}):
\bea
\delta E&=&
\int^{\infty}_0d\omega_1\int^{\infty}_0d\omega_2
\frac{2\omega_2 I(\omega_1,a,L)I(\omega_2,a,L)}{\omega_1^2-\omega_2^2}\biggl[\frac{(\omega_1+\omega_A+\omega_B)\omega_A\omega_B}{(\omega_1+\omega_A)(\omega_1+\omega_B)(\omega_A+\omega_B)}\nn\\&&
+\frac{2\omega_A^2\omega_B^2}{(\omega_1^2-\omega_A^2)(\omega_1^2-\omega_B^2)(e^{2\pi\omega_1/a}-1)}\biggr]
\label{acc-tot-contribution}
\eea
with
\bea
I(\omega_i,a,L)=\frac{3a}{8\pi^2L^2}\bigg[\frac{(1-\frac{1}{2}a^2L^2)\omega_i L}{\mathcal{N}^2(a,L)}\cos(\omega_i D_a)
-\frac{1+a^2L^2+\omega_i^2L^2\mathcal{N}(a,L)}{\mathcal{N}^{5/2}(a,L)}\sin(\omega_i D_a)\bigg],\;\label{I}
\eea
$i=1,2$, $\mathcal{N}(a,L)=1+\frac{1}{4}a^2L^2$, and $D_a=\frac{2}{a}\sinh^{-1}\left({aL\/2}\right)$. Here, we have rescaled the interaction energy in unit of $\alpha(A)\alpha(B)$, the product of the detectors' polarizability  $\alpha(\xi)=\frac{2|\mu^{\xi}_{ge}|^2}{3\omega_{\xi}}$. 

Since $I(\omega_i,a,L)$ approaches zero as $a\rightarrow0$, the interaction energy Eq.~(\ref{acc-tot-contribution}) tends to zero in this limit. This implies that no interaction exists between two cross-polarizable inertial detectors. This result starkly contrasts with outcome of a toy model (in which the detectors are assumed to be in monopole coupling with a scalar field~\cite{Passante14,Cheng22}), where the interaction energy is present for two detectors in both the inertial and acceleration scenarios.

The vanishing interaction between two accelerated cross-polarizable detectors as  $a\rightarrow0$
suggests an intriguing application for distinguishing between inertial and accelerated reference frames. Specifically, the absence of an interaction between two cross-polarizable detectors indicates an inertial frame, while its presence implies an accelerated one. This approach offers an alternative to classical methods, such as checking whether Newton's second law holds, for determining the nature of a reference frame. Additionally, the characteristic of the inter-detector interaction that it emerges in the presence of acceleration but vanishes in the inertial case parallels the behavior of spontaneous excitation in a single detector. Moreover, this interaction does not occur between two static detectors in a thermal bath~\cite{Cheng23}, which sharply contrasts with the FDU effect, where accelerated detectors behave as if they were in a thermal bath at the Unruh temperature.

Note that the inter-detector interaction energy generally depends on the acceleration $a$, the inter-detector separation $L$, and the detectors' transition frequency $\omega_{\xi}$. For simplicity, we next assume that the transition frequencies of the two detectors are identical, denoted by $\omega$. Notably, the inverses of $\omega$ and $a$ introduce two characteristic length scales:  $\lambda=2\pi\omega^{-1}$ and $L_a= a^{-1}$. Then these three length scales $L$, $\lambda$ and $L_a$ define six typical regions: $L\ll\lambda\ll L_a$, $L\ll L_a\ll\lambda$, and $L_a\ll L\ll\lambda$, where the inter-detector separation $L$ is much smaller than the transition wavelength of the detectors $\lambda$;  and $\lambda\ll L\ll L_a$, $\lambda\ll L_a\ll L$, and $L_a\ll\lambda\ll L$, where $L\gg\lambda$. We will examine the impacts of acceleration in these six typical regions. The detailed derivations are provided in Appendix~\ref{Appendix-limitingcases}.

\subsection{Acceleration effects in region $L\ll\lambda$.}

When $L\ll\lambda\ll L_a$ and $L\ll L_a\ll\lambda$, which correspond to very low and moderate acceleration respectively, the inter-detector interaction energy displays a common behavior:
\beq
\delta E\approx
-\frac{9a^2\omega}{512\pi^2L^4}\;,
\label{Interpotential-I-II}
\eeq
which corresponds to an attractive force that scales as $\sim a^2L^{-5}$. Noteworthily, at least one of the two length parameters $L$ and $\lambda$ in these two regions is much smaller than $L_a$.

However, when the acceleration is very large, such that both $L$ and $\lambda$ are well beyond $L_a$, i.e., $L_a\ll L\ll\lambda$, the interaction energy is approximated by
\beq
\delta E\approx-\frac{18\omega^2}{\pi^3a^3L^8}-\frac{9\omega^3}{8\pi^2a^2L^6}\;,
\eeq
where the dominance of either term depends on the subtle relationship among the three characteristic length scales: $L$, $\lambda$ and $L_a$. Specifically, the interaction energy is dominated by the first term when $L_a\ll L\ll\sqrt{\lambda L_a}$:
\beq
\delta E\approx-\frac{18\omega^2}{\pi^3a^3L^8}\;,\label{vdW-III-I}
\eeq
and by the second term when $\sqrt{\lambda L_a}\ll L\ll\lambda$:
\beq
\delta E\approx-\frac{9\omega^3}{8\pi^2a^2L^6}\;.\label{vdW-III-II}
\eeq
Both the two interaction energies correspond to attractive inter-detector interaction forces, with the first scaling as  $\sim a^{-3}L^{-9}$ and the second as $\sim a^{-2}L^{-7}$.

Now some comments are in order.

First, the interaction energy exhibits distinct scaling behaviors $\sim a^{-3}L^{-8}$ and $\sim a^{-2}L^{-6}$ in the two subregions $L_a\ll L\ll\sqrt{\lambda L_a}$ and $\sqrt{\lambda L_a}\ll L\ll\lambda$ of $L_a\ll L\ll\lambda$. Note that  for given values of $a$ and $\lambda$, the former subregion $L_a\ll L\ll\sqrt{\lambda L_a}$ is closer to $L\ll L_a\ll\lambda$ than the latter $\sqrt{\lambda L_a}\ll L\ll\lambda$. These behaviors are notably different from the $\sim a^2L^{-4}$ observed for two detectors at low or moderate acceleration (see Eq.~(\ref{Interpotential-I-II})).

Second, the $L^{-4}$-dependence of the interaction energy in the first two regimes $L\ll\lambda\ll L_a$ and $L\ll L_a\ll\lambda$ and the $L^{-8}$-dependence in $L_a\ll L\ll\sqrt{\lambda L_a}$ are markedly different from the usual dispersive interaction energy between two non-cross-polarizable atoms, which scales as $\sim L^{-6}$ in the region $L\ll\lambda$~\cite{London30}. However, in the farther subregion $\sqrt{\lambda L_a}\ll L\ll\lambda$, the interaction energy's $\sim L^{-6}$ dependence aligns with that of non-cross-polarizable, inertial atoms. Therefore, the new length scale $\sqrt{\lambda L_a}$ in the region $L_a\ll L\ll\lambda$ delineates the point beyond which the interaction energy in the region $L\ll\lambda$ mirrors  the typical $L^{-6}$ behavior seen in  non-cross-polarizable, inertial atoms.

Third, the subregion $\sqrt{\lambda L_a}\ll L\ll\lambda$ exists exclusively for two uniformly accelerated, cross-polarizable detectors; this subregion does not exist if the detectors are inertial.

\subsection{Acceleration effects in region $L\gg\lambda$.}

The interaction energy in the region $\lambda\ll L\ll L_a$, which corresponds to low acceleration, is approximated by
\beq
\delta E\approx-\frac{27a^2}{512\pi^3L^5}\;;\label{Interpotential-IV}
\eeq
while in the region $\lambda\ll L_a\ll L$, which corresponds to a moderate acceleration, it  is given by
\beq
\delta E\approx-\frac{54}{\pi^3a^3L^{10}}\;. \label{Interpotential-V}
\eeq
Both expressions describe an attractive interaction force.

When the acceleration is very large, such that $L_a\ll\lambda\ll L$, the inter-detector interaction energy becomes:
\beq
\delta E\approx-\frac{9\omega^3}{8\pi^2a^2L^6}\;,\label{Interpotential-VI}
\eeq
which again corresponds to an attractive interaction force. Strikingly, this expression of the interaction energy of two detectors with $L\gg\lambda$ coincides with that of two detectors with $\sqrt{\lambda L_a}\ll L\ll\lambda$ (see Eq.~(\ref{vdW-III-II})). This is remarkable because the interaction energy of two detectors with $L\ll\lambda$ and that of two with $L\gg\lambda$ generally display distinctive behaviors, when subjected to low and moderate accelerations, as shown in Eqs.~(\ref{Interpotential-I-II}), (\ref{Interpotential-IV}), and (\ref{Interpotential-V}).

Note that the condition $\sqrt{\lambda L_a}\ll L\ll\lambda$ indicates both $\lambda\gg L_a$ and $L\gg\frac{\lambda}{L}L_a$, which are naturally satisfied within the region $L_a\ll\lambda\ll L$. Consequently, the transition wavelength of the two detectors $\lambda$ and the inter-detector separation $L$ in these two regions, $\sqrt{\lambda L_a}\ll L\ll\lambda$ and $L_a\ll\lambda\ll L$, extend well beyond $L_a$. Once this condition is met, the inter-detector interaction energy displays a universal behavior across both  $L\ll\lambda$ and $L\gg\lambda$ regions.

Additionally, the interaction energy becomes vanishingly small as $a\rightarrow0$ in regions such as $L\ll\lambda\ll L_a$ and $\lambda\ll L\ll L_a$, where both the transition wavelength of the two detectors $\lambda$ and the inter-detector separation $L$ are well within $L_a$. This aligns with the earlier conclusion that no interaction exists between two cross-polarizable inertial detectors. However, the $a\rightarrow0$ inertial limit  does not apply to the other four regions, making the inter-detector interaction in these regions particularly unusual.

\subsection{Repulsive inter-detector interaction.}

So far, our analytical results suggest an interaction force between two cross-polarizable, accelerated detectors that appears to be always attractive. To verify this inference, we next examine the interaction force  $F(a,L)=-\frac{\partial}{\partial L}\delta E(a,L)$ [details are given in Appendix~\ref{Appendix-force}]. Our findings indicate that the force is not necessarily always attractive.

\begin{figure}[H]
\centering
\includegraphics[width=8.0cm]{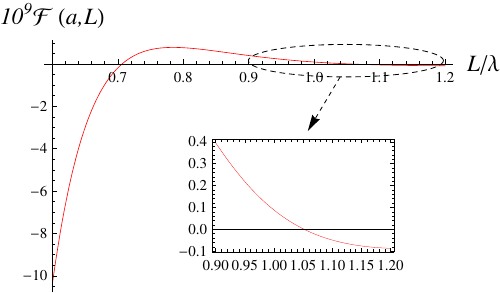}
\caption{\scriptsize Separation-dependence of $\mathscr{F}(a,L)$ for $a\sim0.9\omega$.}
\label{force}
\end{figure}

We resort to numerical computations to determine  the conditions under which a repulsive inter-detector interaction occurs, and the results are displayed in Fig.~\ref{force}. As shown, when  $a\sim\omega$ or equivalently $L_a\sim\lambda$, the value of the inter-detector interaction force $\mathscr{F}(a,L)=(36\pi\omega^8)^{-1}F(a,L)$ oscillates between negative and positive as the inter-detector separation $L$ varies. This indicates that the interaction force can not only be repulsive but can also switch between attractive and repulsive. This behavior contrasts sharply with the usual attractive dispersive interaction force between two inertial atoms, which occurs only when the two atoms are non-cross-polarizable but vanishes when they are cross-polarizable~\cite{Casimir-Polder48}.

\subsection{Duality in acceleration-dependence.}

As evident from Eqs.~(\ref{Interpotential-I-II}), (\ref{vdW-III-II}), (\ref{Interpotential-IV}), and (\ref{Interpotential-VI}), the inter-detector interaction energy for given values of $L$ and $\lambda$ in both regions $L\ll\lambda$ and $L\gg\lambda$ scales as $\sim a^2$ when the acceleration $a$ is very small, and as $a^{-2}$ when $a$ is very large. This behavior reveals a large-small acceleration duality in the acceleration-dependence of the interaction energy.

This large-small acceleration duality suggests that the inter-detector interaction energy becomes vanishingly small in both the small and large acceleration limits. This finding is surprising, as one might expect that a larger acceleration would lead to a stronger interaction, similar to how a higher temperature in a thermal bath generally corresponds to a stronger interatomic interaction. However, in contrast, the inter-detector interaction under our consideration first increases and then decreases as the acceleration grows from very small to very large in both regimes of $L\ll\lambda$ and $L\gg\lambda$.

This characteristic of the inter-detector interaction suggests the existence of an optimal acceleration  $a_c$, which maximizes the interaction strength in each of the two regions, $L\ll\lambda$ and $L\gg\lambda$. For a rough estimation of  $a_c$, one could balance the force that increases with acceleration against the force that decreases with it.  Using this approach, an order of magnitude estimate for $a_c$ in the regime $L\ll\lambda$ yields  $a_c\sim4.43\sqrt{\pi}(\lambda L)^{-1/2}$.

\section{Conclusions}
We discovered an interaction between two Unruh-DeWitt detectors in a vacuum, where one detector is polarizable along the direction of acceleration and the other along the constant, perpendicular inter-detector separation. This interaction emerges only when the detectors are accelerated and vanishes when they become inertial, similar to the spontaneous excitation observed in a single accelerated detector in a vacuum. Notably, this interaction does not occur for two static, cross-polarizable detectors in a thermal bath, which is striking given that the Fulling-Davies-Unruh (FDU) effect predicts that accelerated detectors should behave similarly to static ones in a vacuum.

This novel interaction, which is indispensably dependent on acceleration, exhibits several fascinating and unique properties. First, depending on whether the acceleration is low, moderate, or high, the interaction energy displays distinct separation dependencies in the regions $L\ll\lambda$ and $L\gg\lambda$. These dependencies are markedly different from the typical dispersive interaction energies observed between inertial, non-cross-polarizable atoms. Furthermore, in the case of very large acceleration, where both the inter-detector separation $L$ and the transition wavelength $\lambda$ of the detectors extend well beyond $L_a\equiv a^{-1}$, the interaction energy shows a common behavior across both $L\ll\lambda$ and $L\gg\lambda$. This is in stark contrast to the usual dispersive interaction energy of two inertial, non-cross-polarizable atoms, which exhibits distinctly different behaviors across the entire range of $L\ll\lambda$ and $L\gg\lambda$.

Second, the interaction between two accelerated, cross-polarizable detectors can be either attractive or repulsive, and it may switch between these two behaviors as the inter-detector separation changes, especially when $\lambda\sim L_a$. This behavior stands in sharp contrast to the usual attractive dispersive interaction observed between two inertial, non-cross-polarizable atoms, where the interaction is consistently attractive.

Third, this interaction exhibits a duality in its acceleration-dependence: it behaves as $a^2$ when the acceleration $a$ tends to be zero and as $a^{-2}$ when $a$ becomes extremely large, in both regions of $L\ll\lambda$ and $L\gg\lambda$. This suggests that the inter-detector interaction diminishes in both the low and high acceleration limits, implying the existence of an optimal acceleration at which the interaction is maximized in both regimes.

Finally, it is worth summarizing the similarities and differences between the novel inter-detector interaction we have discovered and the FDU effect. Both effects share a key similarity: they emerge only when the detectors are accelerated and vanish when the detectors are inertial. However, there are notable distinctions between the two. For example, no interaction exists between two static cross-polarizable detectors in a thermal bath, whereas the FDU effect predicts that a single accelerated detector behaves as if it were in a thermal bath. Moreover, the inter-detector interaction we have uncovered exhibits a non-monotonic dependence on acceleration, with an optimal acceleration that maximizes the interaction strength. This behavior contrasts sharply with the monotonic dependence of the detector's response rate in the FDU effect.

\begin{acknowledgments}
This work was supported in part by the NSFC under Grants No. 11690034, No. 12075084, No. 11875172, and No. 12047551, the Zhejiang Provincial Natural Science Foundation of China under Grant No. LY24A050001, and the K. C. Wong Magna Fund in Ningbo University.
\end{acknowledgments}

\appendix

\section{Derivation of Eq.~(10).}\label{Appendix-derivation1}

To calculate the inter-detector interaction energy using the DDC formalism, we must first derive $C^F(\mathrm{x}_A(\tau),\mathrm{x}_B(\tau'))$ and $\chi^F(\mathrm{x}_A(\tau),\mathrm{x}_B(\tau'))$, which are the symmetric correlation function and the linear susceptibility of the field for two arbitrary points along the detectors' trajectories. For this purpose, we start from the quantization  of the field.

In free space, the free part of the electric field operator, $\mathbf{E}^f(\mathrm{x}(\tau))$, can be expressed as~\cite{Cheng23}
\beq
\bm{\mathrm{E}}^f(\mathrm{x}(\tau))=\int d^3\mathbf{k}g_{\mathbf{k}}\sum\limits_{\nu=1}^{2}i\omega_{\mathbf{k}}\bm{\epsilon}(\mathbf{k},\nu)
\left[{a}^f_{\mathbf{k},\nu}(t(\tau))e^{i\mathbf{k}\cdot\mathbf{x}}-{a}^{\dag f}_{\mathbf{k},\nu}(t(\tau))e^{-i\mathbf{k}\cdot\mathbf{x}}\right]\;,
\eeq
where $g_{\mathbf{k}}=[2\omega_{\mathbf{k}}(2\pi)^3]^{-1/2}$, $a^f_{\mathbf{k},\nu}(t(\tau))=a^f_{\mathbf{k},\nu}(t(\tau_0))e^{-i\omega_{\mathbf{k},\nu} [t(\tau)-t(\tau_0)]}$, 
and $\bm{\epsilon}(\mathbf{k},\nu)$ denotes the polarization vectors. 

Combining the above expression and the detectors' trajectories (Eqs.~(1) and (2)) with the definitions of the symmetric correlation function and the linear susceptibility of the field (Eqs.~(5) and (6)), and performing a Lorentz transformation, we obtain
\beq
C^F(\mathrm{x}_A(\tau),\mathrm{x}_B(\tau'))=\frac{1}{6}\int^{\infty}_0d\omega_1\; I(\omega_1,a,L)\left(e^{i\omega_1(\tau-\tau')}+e^{-i\omega_1(\tau-\tau')}\right)\coth(\pi\omega_1/a)
\label{CF-acc}
\eeq
and
\beq
\chi^F(\mathrm{x}_A(\tau),\mathrm{x}_B(\tau'))=\frac{1}{6}\int^{\infty}_0d\omega_1\; I(\omega_1,a,L)\left(e^{-i\omega_1(\tau-\tau')}-e^{i\omega_1(\tau-\tau')}\right)
\label{ChiF-acc}
\eeq
with detailed expression of $I(\omega_1,a,L)$ given in Eq.~(\ref{I}).

Putting Eqs.~(\ref{CF-acc}) and (\ref{ChiF-acc}) above and Eqs. (7) and (9) 
into the two basic formulae describing the contributions of the vacuum fluctuations [vf-contribution] and those of the radiation reaction of the detectors [rr-contribution] (Eqs.~(4) and (8)), and performing the triple integrals with respect to $\tau_1$, $\tau_2$ and $\tau_3$ over an infinite time interval $\Delta\tau=\tau-\tau_0\rightarrow\infty$, we obtain the following vf- and rr-contributions to the inter-detector interaction energy:
\beq
(\delta E)_{vf}=\int^{\infty}_0d\omega_1\int^{\infty}_0d\omega_2\;\frac{I(\omega_1,a,L)I(\omega_2,a,L)\omega_2\omega_A^2\omega_B^2\coth(\pi\omega_1/a)}{(\omega_1^2-\omega_2^2)(\omega_1^2-\omega_A^2)(\omega_1^2-\omega_B^2)}
\label{acc-vf-contribution}
\eeq
and
\bea
(\delta E)_{rr}&=&\int^{\infty}_0d\omega_1\int^{\infty}_0d\omega_2\; I(\omega_1,a,L)I(\omega_2,a,L)
\biggl[\frac{\omega_1\omega_2\omega_A\omega_B^2(\omega_1^2+\omega_2^2-\omega_A^2-\omega_B^2)}
{(\omega_1^2-\omega_A^2)(\omega_2^2-\omega_A^2)(\omega_1^2-\omega_B^2)(\omega_2^2-\omega_B^2)}\nn\\&&
-\frac{\omega_1\omega_2\omega_A\omega_B}{(\omega_A+\omega_B)(\omega_1^2-\omega_B^2)(\omega_2^2-\omega_B^2)}
+\frac{\omega_2\omega_A^2\omega_B^2\coth(\pi\omega_1/a)}{(\omega_1^2-\omega_2^2)(\omega_1^2-\omega_A^2)(\omega_1^2-\omega_B^2)}\biggr]\;.
\label{acc-rr-contribution}
\eea
As mentioned below Eq.~(\ref{I}), we have rescaled the interaction energy in unit of the product of the detectors' polarizability.


Finally, the summation of Eqs.~(\ref{acc-vf-contribution}) and (\ref{acc-rr-contribution}) gives rise to the following total inter-detector interaction energy:
\bea
\delta E&=&
\int^{\infty}_0d\omega_1\int^{\infty}_0d\omega_2\;
\frac{2\omega_2 I(\omega_1,a,L)I(\omega_2,a,L)}{\omega_1^2-\omega_2^2}\biggl[\frac{(\omega_1+\omega_A+\omega_B)\omega_A\omega_B}{(\omega_1+\omega_A)(\omega_1+\omega_B)(\omega_A+\omega_B)}\nn\\&&
+\frac{2\omega_A^2\omega_B^2}{(\omega_1^2-\omega_A^2)(\omega_1^2-\omega_B^2)(e^{2\pi\omega_1/a}-1)}\biggr]\;,
\eea
which is Eq.~(10) in the main text.

\section{Inter-detector interaction energy in limiting cases.}\label{Appendix-limitingcases}

Performing the $\omega_2-$integration in Eq.~(\ref{acc-tot-contribution}) with the contour integration technique and the residue theory, we can simplify Eq.~(\ref{acc-tot-contribution}) to
\bea
\delta E&=&-\frac{36\pi\omega_A\omega_B}{\omega_A+\omega_B}\int^{\infty}_0d\omega_1
\left[\left(A_4\omega_1^4+A_2\omega_1^2+A_0\right)\sin(2\omega_1D_a)+\left(A_3\omega_1^3+A_1\omega_1\right)\cos(2\omega_1D_a)\right]\nn\\&&
\times\left[\frac{(\omega_1+\omega_A+\omega_B)}{(\omega_1+\omega_A)(\omega_1+\omega_B)}+\frac{2\omega_A\omega_B}{\omega_A-\omega_B}
\left(\frac{1}{\omega_1^2-\omega_A^2}-\frac{1}{\omega_1^2-\omega_B^2}\right)\frac{1}{e^{2\pi\omega_1/a}-1}\right]\;,
\label{acc-tot-contribution-1}
\eea
where $A_i\equiv A_i(a,L)$ with $i=0,1,2,3,4$, and
\bea
\left\{
  \begin{array}{ll}
A_0(a,L)=\frac{a^2(1+a^2L^2)^2}{512\pi^4L^4\mathcal{N}^5(a,L)}\;,\\
A_1(a,L)=-\frac{a^2(1-\frac{1}{2}a^2L^2)(1+a^2L^2)}{256\pi^4L^3\mathcal{N}^{9/2}(a,L)}\;,\\
A_2(a,L)=\frac{a^2(1+3a^2L^2-\frac{1}{4}a^4L^4)}{512\pi^4L^2\mathcal{N}^4(a,L)}\;,\\
A_3(a,L)=-\frac{a^2(1-\frac{1}{2}a^2L^2)}{256\pi^4L\mathcal{N}^{7/2}(a,L)}\;,\\
A_4(a,L)=\frac{a^2}{512\pi^4\mathcal{N}^3(a,L)}\;.\\
  \end{array}
\right.
\eea
For arbitrary values of the parameters $a$, $L$, and $\omega_{\xi}$, further evaluation of Eq.~(\ref{acc-tot-contribution-1}) in closed form is rather formidable. However, we can obtain some analytical results in certain limiting cases. 

Before further simplify Eq.~(\ref{acc-tot-contribution-1}) directly, let us note that the integrals in Eq.~(\ref{acc-tot-contribution-1}) can be classified into two groups. The first group consists of integrals without the factor $(e^{2\pi\omega_1/a}-1)^{-1}$, which can be further expressed in terms of some special functions after the $\omega_1$-integration with the help of the contour integration  technique.  The second group is composed of integrals which are characterized by the factor $(e^{2\pi\omega_1/a}-1)^{-1}$,  and these take the form
\beq
I_n=\int^{\infty}_0d\omega_1{\mathfrak{g}_n(\omega_1,D_a)\/(\omega_1^2-\omega_{\xi}^2)(e^{2\pi\omega_1/a}-1)}\;,\label{pole-integral}
\eeq
where $n$ is an integer ranging from 0 to 4, and
\bea
\mathfrak{g}_n(\omega_1,D_a)=\left\{
                         \begin{array}{ll}
                          \omega_1^n\sin({2\omega_1D_a}), n=0,2,4,\\
                           \omega_1^n\cos({2\omega_1D_a}), n=1,3.
                         \end{array}
                       \right.
\eea
We next demonstrate how to deal with these integrals in the limiting cases.

\subsection{
Derivations of Eq.~(12) for $L\ll\lambda_{\xi}\ll L_a$, and Eqs.~(16) and (17). }

When $\lambda_{\xi}\ll L_a$ or equivalently $a\ll\omega_{\xi}$, we first divide the integral in Eq.~(\ref{pole-integral}) into two parts, i.e.,
\beq
I_n=\int^{\omega_{\xi}}_0d\omega_1{\mathfrak{g}_n(\omega_1,D_a)\/(\omega_1^2-\omega_{\xi}^2)(e^{2\pi\omega_1/a}-1)}
+\int^{\infty}_{\omega_{\xi}}d\omega_1{\mathfrak{g}_n(\omega_1,D_a)\/(\omega_1^2-\omega_{\xi}^2)(e^{2\pi\omega_1/a}-1)}\;.
\label{pole-integral-twoparts}
\eeq
Since $a\ll\omega_{\xi}$, it is  easy to  see that the value of the second integral on the right is very small and thus can be neglected. Meanwhile, the upper limit $\omega_{\xi}$ of the first integral could be replaced by infinity. In this way, the first integral and thus $I_n$ can be solved, and $\delta E$ in the limit $\lambda_{\xi}\ll L_a$ is then expressed in terms of some special functions.

Lastly, expanding the results in limiting cases of $L/\lambda_{\xi}\ll1$ with $aL\ll1$, $L/\lambda_{\xi}\gg1$ with $aL\ll1$, and $L/\lambda_{\xi}\gg1$ with $aL\gg 1$, we obtain approximate analytical expressions, Eqs.~(12), (16) and (17), for the inter-detector interaction energy in the three regions $L\ll\lambda_{\xi}\ll L_a$, $\lambda_{\xi}\ll L\ll L_a$, and $\lambda_{\xi}\ll L_a\ll L$. 

\subsection{
Derivations of Eq.~(12) for $L\ll L_a\ll\lambda_{\xi}$, and Eqs.~(13) and (18). }

When $\lambda_{\xi}\gg L_a$ or equivalently $a\gg\omega_{\xi}$, 
we expand the factor $(e^{2\pi\omega_1/a}-1)^{-1}$ in $I_n$ into series, transform $\omega_1\pm\omega_{\xi}$ into a new variable $t$, and then arrive at
\bea
I_n&=&\frac{1}{2\omega_{\xi}}\sum^{\infty}_{m=1}e^{-2\pi m\omega_{\xi}/a}\int^{\infty}_{-\omega_{\xi}}dt\;\frac{\mathfrak{g}_n(t+\omega_{\xi},D_a)e^{-2\pi mt/a}}{t}\nn\\&&
-\frac{1}{2\omega_{\xi}}\sum^{\infty}_{m=1}e^{2\pi m\omega_{\xi}/a}\int^{\infty}_{\omega_{\xi}}dt\;\frac{\mathfrak{g}_n(t-\omega_{\xi},D_a)e^{2\pi mt/a}}{t}\;.
\label{second-high-acceleration}
\eea
Now considering that $\omega_{\xi}/a\ll1$, we can replace the infinite summation over $m$ in the above equation by an integration over $y(=m\omega_{\xi}/a)$, which can be evaluated directly. In this way, the $t-$intergral in the above equation and thus the total interaction energy $\delta E$ can be expressed in terms of some special functions.

Next, expanding the results 
in limiting cases of $L/\lambda_{\xi}\ll1$ with $aL\ll1$, $L/\lambda_{\xi}\ll1$ with $a L\gg1$, and $L/\lambda_{\xi}\gg1$ with $aL\gg1$,  we obtain the approximate analytical results for the inter-detector interaction energy, Eqs.~(12), 
(13) and (18), in regions $L\ll L_a\ll\lambda_{\xi}$, $L_a\ll L\ll\lambda_{\xi}$, and $L_a\ll\lambda_{\xi}\ll L$, respectively.

Here it is worth emphasizing that the transition frequencies of the two detectors are set to be identical in the discussions of the main text for simplicity, i.e., $\omega_{\xi}=\omega$. Additionally, we have checked all the approximate analytical results in the six typical regions with numerical computations, and good agreements have been achieved.

\section{Details about the inter-detector interaction force.}\label{Appendix-force}

To discuss the interaction force numerically, we first resort to the relation $F(a,L)=-\frac{\partial}{\partial L}\delta E(a,L)$ and get
\beq
F(a,L)=36\pi\omega^8\mathcal{F}(a,L)\;,\label{cross-pol-acc-force}
\eeq
where $\mathcal{F}(a,L)$ is a dimensionless function given by
\bea
\mathcal{F}(a,L)&=&\int^{\infty}_0dx\frac{B_0-B_1x-B_2x^2+B_3x^3+B_4x^4-B_5x^5}{(x^2+1)^2}e^{-2\omega D_a x}\nn\\&&
+\sum_{i=0,2,4}\int^{\infty}_0dx\frac{2B_i\sin(2\omega D_a x)}{(x^2-1)^2(e^{2\pi\omega x/a}-1)}\nn\\&&
+\sum_{i=1,3,5}\int^{\infty}_0dx\frac{2B_i\cos(2\omega D_a x)}{(x^2-1)^2(e^{2\pi\omega x/a}-1)}
\label{mathcalf}
\eea
with $B_i \equiv B_i(a,L)$ and
\bea
\left\{
  \begin{array}{ll}
    B_0(a,L)=-\frac{a^2(1+a^2L^2)(8+7a^2L^2+5a^4L^4)}{1024\pi^4\omega^7L^5\mathcal{N}^6(a,L)}\;,\\
    B_1(a,L)=\frac{a^2(16+22a^2L^2+11a^4L^4-4a^6L^6)}{1024\pi^4\omega^6L^4\mathcal{N}^{11/2}(a,L)}\;,\\
    B_2(a,L)=\frac{a^2(16-4a^2L^2-68a^4L^4+3a^6L^6)}{4096\pi^4\omega^6L^3\mathcal{N}^5(a,L)}\;,\\
    B_3(a,L)=-\frac{a^2(4+11a^2L^2-2a^4L^4)}{512\pi^4\omega^4L^2\mathcal{N}^{9/2}(a,L)}\;,\\
    B_4(a,L)=\frac{a^2(8-7a^2L^2)}{1024\pi^4\omega^3L\mathcal{N}^4(a,L)}\;,\\
    B_5(a,L)=\frac{a^2}{256\pi^4\omega^2\mathcal{N}^{7/2}(a,L)}\;.\\
  \end{array}
\right.
\eea
Next, we can numerically explore the separation- and acceleration-dependencies of the interaction force by excluding the disposable poles. That's how the numerical results in Fig.~1 are obtained.

\end{document}